\documentstyle[12pt]{article}
\newcommand {\be}{\begin{equation}}
\newcommand {\ee}{\end{equation}}
\newcommand {\bea}{\begin{eqnarray}}
\newcommand {\eea}{\end{eqnarray}}
\newcommand {\pder}[2]{\frac{\partial #1}{\partial #2}}

\newcommand {\vett}[1] {{\mathbf #1}}
\newcommand{\bsym}[1]{\hbox{\rm \boldmath{$#1$}}}
\newcommand{\symvar}[1]{\hbox{\rm \boldmath{$#1$}}(\vett r,t)}
\newcommand{\vecvar}[1]{{\mathbf #1}({\mathbf r},t)}
\newcommand{\scavar}[1]{#1({\mathbf r},t)}
\newcommand{\opevar}[1]{\hat{#1}({\mathbf r},t)}
\newcommand{\psim}{\langle\opevar{\psi}\rangle}
\renewcommand{\Re}{\mathrm{Re}}
\renewcommand{\Im}{\mathrm{Im}}
\renewcommand{\(}{\left(}
\renewcommand{\)}{\right)}
\renewcommand{\[}{\left[}
\renewcommand{\]}{\right]}

\begin{document}

\title{\bf Time-Dependent Density-Functional Theory for Superfluids}

\author{M.~L. Chiofalo 
and M.~P. Tosi
\\
{\small \it INFM and Classe di Scienze,}\\
{\small \it  Scuola Normale Superiore, 
I-56126 Pisa, Italy}\\
}

\date{}
\maketitle


\begin{abstract}
A density-functional theory is established for inhomogeneous
superfluids at finite temperature, 
subject to time-dependent external fields in 
isothermal conditions. 
After outlining parallelisms between a neutral
superfluid and a charged superconductor, Hohenberg-Kohn-Sham-type
theorems are proved for gauge-invariant densities and a set of
Bogolubov-Popov equations including exchange and correlation is set
up. Earlier results applying in the linear response regime are
recovered. 

\end{abstract}

\newpage
Experiments on confined Bose-condensed gases have revealed a
rich variety of dynamical behaviours.
These include elementary excitations of low-lying shape deformation 
modes \cite{exp1}, propagation of sound
waves in both the condensate and the thermal cloud \cite{exp5}, 
Josephson-like oscillations in double condensates \cite{exp6}, 
creation of vortices \cite{vortex}, phase dynamics from 
various atom-laser configurations \cite{atomlas} and 
Bloch oscillations of a
condensate in an optical lattice \cite{ak}. 

While mean-field
theories suffice in most cases to describe 
the observed behaviours, 
condensates can now be created where effects beyond mean-field can 
be explored, by tuning almost at will the scattering length and 
hence the condensate self-interaction energy \cite{Rb85}
or else by approaching the critical temperature for
Bose-Einstein condensation. 
Along these lines a Time-Dependent Density-Functional Theory 
(TD-DFT) for superfluids
is a suitable framework to treat their dynamics with
inclusion of exchange and correlation. 

For a many-electron system in the normal state the foundations of the
theory come from a set of theorems by Runge and Gross \cite{RG,GK},
which have been extended to superconductors at zero
temperature by Wacker, K\"ummel and Gross (WKG) 
\cite{WKG}. In applications to normal
electron systems in the linear response regime, 
the limitations to low-frequency phenomena 
have been conceptually overcome by 
Vignale and Kohn \cite{VK}. Their
current-density formulation of TD-DFT
embodies plasmon dispersion and damping as well as transverse-current
fluctuations, allows a unified treatment of the damping
of collective excitations from 
the Landau and mode-coupling mechanisms
and yields microscopic generalized-hydrodynamic equations
\cite{VK,VUC}.

A similar scheme has already been developed for the dynamic linear
response of superfluids \cite{PHB}, extending to inhomogeneous 
systems and to finite-frequency phenomena Landau's hydrodynamic
equations in the two-fluid model. The present Letter concerns the
underlying foundations of TD-DFT for superfluids. The proof of the
relevant Hohenberg-Kohn-Sham-type theorems 
parallels the WKG derivation for superconductors and
one of our  results is a 
``dictionary'' which translates vector and scalar
potentials, Maxwell equations and the like from a charged to 
a neutral fluid characterized by spontaneous symmetry breaking.  
We also allow for
finite temperature in isothermal
conditions. This preludes to a specific choice of the 
reference system for the TD-DFT mapping, which is described by 
a set of Bogolubov-Popov equations including the non-condensate 
density.

\medskip\noindent{\it Introductory material.}
The dynamics of a system of interacting spinless bosons 
confined in a static potential and evolving 
from an initial equilibrium state at 
time $t_0$ is driven by the Hamiltonian 
\be
\hat{H}(t)=
\hat{T}_{\vett A}(t)+\hat{V}_V(t)+\hat{S}_{\eta}(t)+\hat{W}(t)\; .
\label{H}
\ee
The system is subject to an external vector potential
$\vett A$ and to  
scalar fields $V$ and $\eta$. 
We have 
$
\hat{T}_{\vett A}=-(1/2m)
\int d\vett r\opevar{\psi^\dagger}\[
{\hbar\bsym{\nabla}}-i\vecvar{A}\]^2\opevar{\psi}
$,
$
\hat{V}_V=\int d{\vett r} \scavar{V}\opevar{\psi^\dagger}
\opevar{\psi^{}}
$
and
$\hat{S}_{\eta}=
\int d\vett r\[\scavar{\eta}\opevar{\psi^\dagger}
+\scavar{\eta^*}\opevar{\psi}\]$ 
\cite{HM}.  
$\hat{W}(t)$ is the interaction term, given in terms of the 
field operators by 
$
\hat{W}(t)=(1/2)\int d\vett r_1 d\vett r_2
\hat{\psi^\dagger}(\vett r_1, t) 
\hat{\psi^\dagger}(\vett r_2, t) w(\vett r_1, \vett r_2)
\hat{\psi}(\vett r_2, t) 
\hat{\psi}(\vett r_1, t) 
$.

$\vett A$
couples to the total current density and
$V$ to the total particle
density. After writing the ensemble average of the field operator 
as $\langle\opevar{\psi}\rangle=
\sqrt{\scavar{n_c}}\exp[i\scavar{\varphi}]$, we see that  
the symmetry-breaking source field $\eta$ drives 
both the density of condensate $\scavar{n_c}$
and its phase $\scavar{\varphi}$; the latter 
determines the irrotational 
part ${\vett v}_s(\vett r,t)$ of the velocity field through
${\vett v}_s(\vett r,t)=(\hbar/m)\bsym{\nabla}\scavar{\varphi}$. 
In the linear regime
$\hat{S}_{\eta}$ we can write 
$\hat{S}_{\eta}=
\int d{\vett r} [
\symvar{\lambda}\cdot\opevar{\delta{\vett v}_s}+\scavar{\alpha}
\opevar{\delta n_c}]$
in terms of the condensate-density 
operator $\delta\hat{n_c}(\vett r,t)=
2\Re[\langle\hat{\psi}(\vett r,t_0)\rangle
\opevar{\delta\psi^{\dagger}}]$ and of the irrotational-flow 
operator 
$\opevar{\delta{\vett v}_s}=
(\hbar/m)\bsym{\nabla}\delta\hat{\varphi}$ with  
$\delta\opevar{\varphi}=-\Im 
[\opevar{\delta\psi^{\dagger}}/
\langle\hat{\psi}^\dagger(\vett r,t_0)\rangle]$.
The (real) fields $\alpha$ 
and $\bsym{\lambda}$ are then related to $\eta$ by  
$\alpha (\vett r,t)=[n_c(\vett r,t_0)]^{-1}
\Re[\langle\hat{\psi}^\dagger(\vett r,t_0)\rangle
\eta(\vett r, t)]$
and $\bsym{\nabla}\cdot\bsym{\lambda}(\vett r,t)
=-2m\Im [\langle\hat{\psi}^\dagger(\vett r,t_0)
\rangle\eta(\vett r, t)]$
(see \cite{HM}).

The quantity needed to deal with time-dependent phenomena 
in DFT is the quantal action \cite{RG}. 
According to WKG, this is
\be
Q\equiv\int_{t_0}^{\tau}dt < \frac{i\hbar}{2}\int
d\vett r \[\opevar{\psi^\dagger}\pder{\opevar{\psi}}{t}
-\pder{\opevar{\psi^\dagger}}{t}\opevar{\psi}
\]-\hat{H}>\; .
\label{action}
\ee
Following the well-known DFT 
argument, we shall prove below that the potentials are in 
one-to-one
correspondence with appropriate gauge-invariant 
densities (Theorem I); that  
the action functional can be written in terms of these densities 
(Theorem II); and that 
a practical scheme can be given to map the interacting system into a
non-interacting one driven by effective potentials 
which include exchange and correlation (Theorem III).

The action (\ref{action})
is invariant under the gauge transformation
$\vecvar{A}\rightarrow \vecvar{A}+{\hbar}{\bsym{\nabla}}
\scavar{\Lambda}
$
and
$
\scavar{V}\rightarrow \scavar{V}-\hbar
\partial{\scavar{\Lambda}}/\partial {t}
$,
where $\scavar{\Lambda}$ is a real scalar function such that 
$\Lambda(\vett r,t_0)=0$ $mod\; (2\pi)$. 
The source function $\scavar{\eta}$ 
transforms into $\scavar{\eta}\exp[i\scavar{\Lambda}]$, so
that $\hat{S}_{\eta}$ is gauge-invariant. 
The operators transform
according to 
\be
\opevar{\psi}\rightarrow
\opevar{\psi}\exp[i\scavar{\Lambda}]\; \label{gpsi}
\ee
and
\be
\opevar{\vett j}\rightarrow \opevar{\vett j}+
(\hbar/m)\opevar{\psi^\dagger}\opevar{\psi}\bsym{\nabla}
\scavar{\Lambda}\label{gop}\; ,
\ee
with $\hat{\vett j}$ the paramagnetic current-density operator. 
We choose the following gauge-invariant densities: 
the current density 
$\vecvar{j}=\langle\vecvar{\hat{j}}\rangle
-\scavar{n}\vecvar{A}/m$, the condensate density
$\scavar{n_c}=|\langle\opevar{\psi^\dagger}\rangle|^2$
and the velocity field
${\vett v}_s(\vett r,t)=(\hbar/m)\bsym{\nabla}\scavar{\varphi}-
\vecvar{A}/m$. The total 
density $\scavar{n}$ is not an independent quantity, 
since it is   
determined by  the continuity equation. 
The pair of 
physical quantities $\scavar{n_c}$ and $\vecvar{v_s}$
can be replaced by the gauge-invariant order parameter 
$\scavar{\Phi}=\psim\exp[-(i/\hbar)\int_{t_0}^t dt'V(\vett r,t')]$
: we shall exploit this fact whenever convenient. Finally, 
we stress that $\vett v_s$ is the
{\it irrotational} part of the velocity field.

\medskip\noindent{\it The role of $\vett A$ and $V$.}
We have already discussed the role of the source field 
$\eta$, which is characteristic of a neutral superfluid. Before 
proceeding to prove the TD-DFT theorems we pause 
to discuss the physical import of the fields $\vett A$ 
and $V$ entering the Hamiltonian (\ref{H}). Their meaning is obvious
for a superconductor, but needs elaboration for a superfluid. 
We use as a guideline for this purpose the linearized
two-fluid model below threshold for vortex formation. 

We consider first the vector  potential $\vett A$.
As pointed out by Baym while discussing the rotating-bucket 
experiment \cite{baym}, what may create transverse 
currents in a superfluid 
is a spoon-stirring mechanism. Let $\bsym 
\omega$ be the stirring angular velocity, with magnitude 
below the threshold $\omega_c$ for vortex formation. If
$\vett L= m\vett r\times\vett j$ is the angular momentum, 
the corresponding Hamiltonian term can be written as
$\int d\vett r\; \bsym\omega\cdot \vett L=\int d\vett r\; 
\vett j\cdot 
(\bsym\omega\times\vett r)$. Comparing with the minimal coupling 
form $\vett j\cdot\vett A$ in the Hamiltonian (\ref{H}), 
the component of $\vett A$ parallel to 
$\vett j$ is $\vett A=m\bsym\omega\times\vett r$, namely with 
$m$ times the rigid-body rotational velocity of the fluid. 
Since that part of the fluid which can
respond to a transverse probe is by definition 
the normal-fluid component, $\bsym\omega\times\vett r$ is
the normal-fluid velocity $\vett v_n$ and thus  
$\vett A=m{\vett v}_n$. This result remains true for a
non-rotating fluid, as demonstrated by Hohenberg and Martin 
by means of a Galileian transformation \cite{HM}. 

Let us turn to the scalar potential $V$. The quantity 
$V+\hbar
\partial\varphi/\partial t$, with $\varphi$ being the phase of 
the condensate, is gauge-invariant. Therefore,  
writing the equations in the gauge in which the scalar potential
vanishes corresponds to a Galileian transformation to a reference
frame moving with velocity $\vett v_s$. This fact 
will be used in the proof of Theorem I below.

From the above arguments regarding the potentials 
$\vett A$ and $V$, it follows that 
$\partial (\vett v_n-\vett v_s)/\partial t=\partial
\vett A/\partial t+\bsym\nabla V/m$ and therefore 
is gauge-invariant. In fact, 
in the two-fluid model  
(with $n=\rho_s+\rho_n$,
$\rho_s$ and $\rho_n$ being the super- and normal-fluid densities) 
the gauge-invariant current density is
${\vett j}_r=\rho_s\vett v_s+\rho_n\vett v_n-n\vett v_n=\rho_s(\vett
v_s-\vett v_n)$. This is the current as seen in a reference 
frame which moves with the normal-fluid component and determining
one of the driving forces in the Landau-Khalatnikov 
equations \cite{khala}. 

We conclude by remarking that a parallel can be made between 
the two-fluid equations for neutral superfluids
and Maxwell's equations for charged superconductors. 
From the above analysis it turns out 
that  the equation
$\bsym\nabla\times(\vett E+c^{-1}\partial\vett A/\partial
t)=0$ or else $\vett E+c^{-1}\partial\vett A/\partial
t=-\bsym\nabla V$ is just the condition for irrotational 
flow. The ``electric field'' $\vett E$ is identified with 
$(m/\rho_s)\partial \vett j_r/\partial t$. As expected, 
the Maxwell equation for $\bsym\nabla\times\vett B$
expresses the relation of continuity between particle and
current densities.

After this excursus we return to the basic theorems of TD-DFT for neutral superfluids.

\medskip\noindent{\it Theorem I.} It states 
that the densities $\{d\}\equiv\{\vecvar{j},\scavar{\Phi}\}$ 
are uniquely related to the
potentials $\{p\}\equiv\{\vecvar{A},\scavar{V},\scavar{\eta}\}$.
One has to show that two sets of 
potentials $\{p\}$ and $\{p'\}$, which differ 
by more than a gauge   
transformation and can 
be expanded in Taylor series around $t_0$, determine 
two different sets of densities $\{d\}$ and 
$\{d'\}$ evolving from a common 
initial equilibrium state.
While the statement is trivially true at time $t_0$,
it is sufficient to prove it at some time $t$ infinitesimally 
later than $t_0$ by relating the 
coefficients of the Taylor series for the densities 
to those for the potentials \cite{RG}. 

We thus consider the Heisenberg equations of motion for the 
densities. 
The potentials contribute to
the equation for the induced current 
${j_\alpha}$ with terms including 
$n\nabla_\alpha V$, $A_\beta\nabla_\beta j_\alpha$,
$j_\beta\nabla_\alpha A_\beta$, 
$j_\alpha\nabla_\beta A_\beta$,
and $(\hbar n/m)A_\beta\nabla_\alpha A_\beta$.
It is evident that the proof will be easier in reference frame moving
with velocity ${\bf v}_s$: in this gauge, as already remarked, the
scalar potential vanishes and ${\vett j}$ and ${\vett A}$ are both 
transverse, so that all the 
above terms vanish. We proceed within this gauge, 
signalled henceforth by a tilde on the potentials.

Since the two sets of 
potentials are different, their Taylor-expansion coefficients
must differ at some order, say 
$l$ for $\vett{\tilde A}$ and $\vett{\tilde A}'$ and  
$l'$ for $\tilde{\eta}$ and $\tilde{\eta}'$. 
It is then sufficient to show, for the lower among
$l$ and $l'$, that different
coefficients in the expansion of the potentials 
imply different coefficients in the expansion of the densities 
\cite{RG}\nocite{GK}-\cite{WKG}.
In the case $l<l'$ we have 
\be
\frac{\partial^l}{\partial t^l}\[\vecvar{j}-\vecvar{j'}\]_{t=t_0}=
\frac{n(\vett r,t_0)}{m}
\frac{\partial^l}{\partial t^l}\[\vecvar{\tilde{A}}-
\vecvar{\tilde{A}'}\]_{t=t_0}\; ,
\label{Ij}
\ee
while in the case $l>l'$ we have
\be
i\hbar\frac{\partial^{l'+1}}{\partial t^{l'+1}}
\[\scavar{\Phi}-\scavar{\Phi'}\]_{t=t_0}=
\frac{\partial^{l'}}
{\partial t^{l'}}\[\scavar{\tilde{\eta}}-
\scavar{\tilde{\eta}'}\]_{t=t_0}\; .
\label{Ivs}
\ee
As a consequence of eqs. (\ref{Ij}) and (\ref{Ivs}), the set of
densities $\{d\}$ will differ from $\{d'\}$ at times infinitesimally
later than $t_0$. Hence they are different. This proves 
Theorem I.

The conclusion thus is that in a superfluid 
the potentials are unique functionals of the 
densities. Since from the Heisenberg
equations of motion the field operators are functionals of
the potentials, we may state that the ensemble expectation  
value of any gauge-invariant operator is a
unique functional of the chosen set of densities.

\medskip\noindent{\it Theorem II.}
It states that (i) the action $Q$ in given external potentials 
can be  expressed as a unique functional 
$Q^0\[\{d\} \]$ of the  densities $\{d\}$
where the superscript $0$ indicates the external
potentials; and (ii)  
$Q^0\[\{d\} \]$ is stationary with respect to the actual  
densities $\{ d^0\}$ of the interacting system.
The proof precisely parallels that given by WKG for 
superconductors, once their complex gap function 
$\Delta(\vett r,t)$ is replaced by $\scavar{\Phi}$
or by the subset $\{\scavar{n_c},
{\vett v}_s(\vett r,t)\}$.

The functional is given by 
\be
{Q^0}\[\{d\} \]=
R\[\{d\} \]-
W\[\{d\} \]-{P^0}\[\{d\} \]-{S^0}\[\{d\} \]
\label{action0}
\ee
where 
\be
R\[\{d\} \]\equiv(1/2)\int_{t_0}^t dt'\int d\vett r\langle
\hat{\psi}^\dagger\[\{d\} \]\[\(i\hbar\partial/\partial t'\)-
(\hbar^2/2m)\nabla^2\]\hat{\psi}^{}\[\{d\} \]\rangle+c.c.
\ee
and $W\[\{d\} \]\equiv
\int_{t_0}^t dt'\langle \hat{W}\[\{d\} \](t')
\rangle$ are its universal parts, while
\bea
{P^0}\[\{d\} \]&\equiv&\int_{t_0}^t dt'\int d\vett r
\[
\({V^0}(\vett r,t')+\frac{1}{2m}
{{A^0}^2}(\vett r,t')\){n\[\{{d(\vett r,t')}\} \]}
+\right.\nonumber\\
&+&\left.{\vett A}^0(\vett r,t')\cdot\({\vett j}(\vett r,t')-
{n\[\{{d(\vett r,t')}\} \]}
{\vett A}\[\{{d(\vett r,t')}\} \]/m\)
\]
\eea
and
\be
{S^0}\[\{d\} \]\equiv\int_{t_0}^t dt'\int d\vett r
 \[\eta_g^0(\vett r, t'){\Phi^*(\vett r,t')}+
{\eta_g^0}^*(\vett r, t'){\Phi(\vett r,t')}\]\; ,
\ee
depend on the external potentials.
The gauge has been chosen so that 
the functional ${V}\[\{d\} \]$ equals
the external 
scalar potential $\scavar{V^0}$ and
$\eta_g^0$ is defined by  
$\eta_g^0\equiv\eta^0\exp[-(i/\hbar)
\int_{t_0}^t dt'V^0(\vett r,t')]$.

Along with the basic idea underlying DFT, Theorem II 
admits a map
of the densities in the real system onto those of a
reference system subject to appropriate
potentials. 
This map is proven in Theorem III, which
defines the so-called Kohn-Sham scheme needed to
implement TD-DFT. 

\medskip\noindent{\it Theorem III.}
It states that there exist unique reference-potential functionals 
$\{p^R\[\{d\}\]\}$ such that the densities $\{d^R\} $
calculated within the chosen reference system 
coincide with the densities $\{d^0\}$ of the real
interacting system. 

Following again WKG, we 
first define the action functional 
${Q^R}\[\{d\} \]\equiv
R^R\[\{d\} \]-P^R\[\{d\} \]-S^R\[\{d\} \]$ 
for the reference system as
in eq. (\ref{action0}) for its
interacting analogue. The functional 
$Q^0\[\{d\} \]$ is written as
\be
Q^0\[\{d\} \]=R^R\[\{d\} \]-P^0\[\{d\} \]-
S^0\[\{d\} \]-Q_{xc}\[\{d\} \]\; ,
\ee 
thereby defining the exchange-correlation functional 
$Q_{xc}\[\{d\}\]$. We can now
exploite the stationarity of both 
${Q^R}\[\{d\} \]$ and $Q^0\[\{d\} \]$ 
to obtain a  set of equations relating
the potentials $\{p^R\[\{d\}\]\}$ 
to the original external potentials 
$\{p^0\}$.

The resulting equations, in addition to 
$V^R\[\{\scavar{d^0}\}\]=\scavar{V^0}$ are as follows:
\be
{\[{\delta{P^R}\[\{d\} \]\over\delta\vecvar{j}}\]}
_{\{d^0\}}=
{\[{\delta{P^0}\[\{d\} \]\over\delta\vecvar{j}}\]}
_{\{d^0\}}
+{\[{\delta{Q_{xc}}\[\{d\} \]\over\delta\vecvar{j}}\]}
_{\{d^0\}}
\label{2x2j}
\ee
and
\be
\scavar{\eta_g^R}+
{\[{\delta{P^R}\[\{d\} \]\over\delta\scavar{\Phi^*}}\]}
_{\{d^0\}}
=\scavar{\eta_g^0}+
{\[{\delta{P^0}\[\{d\} \]\over\delta\scavar{\Phi^*}}\]}
_{\{d^0\}}+
{\[{\delta{Q_{xc}}\[\{d\} \]\over\delta\scavar{\Phi^*}}\]}
_{\{d^0\}}\; 
\label{2x2}
\ee
with its complex conjugate. 
These equations define the effective exchange-correlation 
potentials.

We conclude this discussion by noticing that eqs. 
(\ref{2x2j}) and (\ref{2x2}) have been derived in a previous
paper \cite{PHB} within a linear-response formulation of TD-DFT
for superfluids. In brief, by writing  the microscopic
equation of motion for the order parameter in terms of 
the condensate self-energy we proved that 
the matrix expressing the
linear response of $\scavar{n_c}$ and ${\vett v}_s(\vett r,t)$ 
to the symmetry-breaking field $\eta$ explicitly has the
Hohenberg-Kohn-Sham structure.

\medskip
\noindent{\it Reference system.} A suitable choice of
the reference system for a superfluid at finite temperature 
is provided by the gapless 
Bogolubov-Popov approximation. This accounts for  
the thermally excited non-condensate cloud and 
satisfies the Hugenholtz-Pines theorem \cite{HM}.
In this approximation 
the densities can be written as \cite{griffin}
\be
\scavar{n_c}=|\scavar\Phi|^2,
\ee
\be
{\vett v}_s(\vett r,t)=
({\hbar/m})\bsym{\nabla} \scavar{\varphi}-{{\vett A}^R
(\vett r, t)/m}
\label{popcond}
\ee
and
\be
\vecvar{j}=\vecvar{j_c}+\vecvar{\tilde{j}}-{\scavar{n}}
{\vett A}^R(\vett r, t)/m\; .
\ee
Here
$\vecvar{j_c}\equiv \scavar{n_c}\vecvar{v_s}$ is the condensate current
and
\be
\vecvar{\tilde{j}}= \frac{1}{2im}\sum_n\[
N_n\scavar{U_n}\bsym{\nabla} \scavar{U_n^*}+\(N_n+1\)
\scavar{V_n}\bsym{\nabla}\scavar{V_n^*}
-c.c.\]
\label{popcurr}
\ee
is the current carried by the non-condensate. In eq. (\ref{popcurr})
$U_n$ and $V_n$ are the Bogolubov functions and 
$N_n=\[\exp(E_n/k_BT)-1\]^{-1}$ is the boson thermal factor, with
$E_n$ being the energy eigenvalues in the 
Bogolubov-Popov equations (see
below) at the initial time $t_0$. Finally, 
$\scavar{n}=\scavar{n_c}+\scavar{\tilde{n}}$, with 
\be
\scavar{\tilde{n}}=\sum_n \[N_n\(|\scavar{U_n}|^2+
|\scavar{V_n}|^2\)+
|\scavar{V_n}|^2\]
\label{popden}
\ee 
being the non-condensate density. The anomalous density 
is given by $\langle\opevar{\psi}\opevar{\psi}\rangle=
\scavar{\Phi^2}+\sum_n \(2N_n+1\)\scavar{U_n}\scavar{V^*_n}$.

In eq. (\ref{popcond}) the condensate wave
function $\scavar{\Phi}$ satisfies the Schr\"odinger
equation 
\cite{fetter}
\be
i\hbar\pder{\scavar{\Phi}}{t}={\mathcal L}^R
\scavar{\Phi}+\scavar{\eta_g^R}\; ,
\label{gp}
\ee
where ${\mathcal L}^R\equiv-({2m})^{-1}
\[\hbar\bsym{\nabla}-i\vett A^R(\vett r,t)\]^2+\scavar{V^R}$ and
$\scavar{V^R}=\scavar{V^0}+2\int d\vett r' w(\vett r-\vett
r')n(\vett r',t)+\scavar{V_{xc}}$, with $\scavar{V_{xc}}$ being 
determined from eq. (\ref{2x2j}). 
The gauge-invariant reference source field is
$\scavar{\eta_g^R}=\scavar{\eta_g^0}-\Phi (\vett r,t)
\int d\vett r' w(\vett r-\vett
r')|\Phi(\vett r',t)|^2+\scavar{\eta_{xc}}=  
\scavar{\eta_g^0}+
\delta Q_{xc}/\delta\Phi^*$ (see eq. (\ref{2x2})).
In the special case of a point-contact interaction 
eq. (\ref{gp}) becomes the well-known
Gross-Pitaevskii equation.

The Bogolubov functions ${U_n}$ and 
${V_n}$ satisfy the single-particle coupled equations 
\be
i\hbar\frac{\partial}{\partial t}
\(\begin{array}{c}\scavar{U_n}\\\scavar{V_n}\\\end{array}\)=
\(\begin{array}{cc}
{\mathcal L}^R
&-\int d\vett r' w(\vett r-\vett
r')\Phi^{*2}(\vett r',t) \\
-\int d\vett r' w(\vett r-\vett
r')\Phi^{2}(\vett r',t) 
&-{\mathcal L}^R \\
\end{array}\)
\(\begin{array}{c}\scavar{U_n}\\\scavar{V_n}\\\end{array}\)
\;. 
\label{bogo}
\ee
We point out that, as a result of imposing gauge invariance, 
the reference system in eq. (\ref{gp}) is the same as
that in eq. (\ref{bogo}). 

In summary, we have demonstrated the basic Hohenberg-Kohn-Sham-type
theorems underlying
TD-DFT for inhomogeneous neutral superfluids at finite temperature
below threshold for vortex formation and proposed an implementation
based on a reference system described by the Bogolubov-Popov theory.
We have also explicitly pointed out similarities
and differences with respect 
to charged superconductors as treated by Wacker {\it et al.} 
\cite{WKG} and briefly remarked on the linear-response limit as 
treated by Chiofalo {\it et al.} \cite{PHB}. 
A final comment is in order. 
For super-critical rotational velocities 
quantized vortices will appear in the superfluid:
at that point 
the velocity field ${\vett v}_s$ ceases to be 
irrotational and acquires a regular contribution ${\vett v}_r$
describing the velocity of each point of a vortex line  
as well as a singular contribution due to the quantized structure 
of the vortex line \cite{NP}.
The regular term 
leads to the well-known Magnus force on a vortex line 
\cite{wexler} and to  friction forces between 
superfluid and normal-fluid components, which are proportional to 
${\vett v}_r-{\vett v}_n$ \cite{NP}. Therefore, in order to 
account for vortices the present TD-DFT approach will need 
extension to include one additional external field and one 
additional density variable. We hope to return to this problem 
in the near future.

\end{document}